\documentclass{pasa}

\usepackage{graphicx}
\usepackage{graphicx,url,amssymb,amsmath,rotating,color,units,wasysym,epsfig,multirow,epstopdf,subcaption}
\newcommand{\ks}{\kappa^2}

\title[An introduction to Bayesian inference in gravitational-wave astronomy]{An introduction to Bayesian inference in gravitational-wave astronomy: parameter estimation, model selection, and hierarchical models}

\author{Eric Thrane$^1,^2,$\thanks{eric.thrane@monash.edu} ~and Colm Talbot$^1,^2,$\thanks{colm.talbot@monash.edu}
\affil{$^1$ Centre for Astrophysics, School of Physics and Astronomy, Monash University, VIC 3800, Australia}%
\affil{$^2$OzGrav: The ARC Centre of Excellence for Gravitational-Wave Discovery, Clayton, VIC 3800, Australia}
}%

\jid{PASA}
\doi{10.1017/pas.\the\year.xxx}
\jyear{\the\year}

\usepackage{aas_macros}
\usepackage{hyperref} 
\hypersetup{colorlinks,citecolor=blue,linkcolor=blue,urlcolor=blue}

\hypersetup{draft}

\begin{document}

\begin{frontmatter}
\maketitle

\begin{abstract}
This is an introduction to Bayesian inference with a focus on hierarchical models and hyper-parameters.
We write primarily for an audience of Bayesian novices, but we hope to provide useful insights for seasoned veterans as well.
Examples are drawn from gravitational-wave astronomy, though we endeavor for the presentation to be understandable to a broader audience.
We begin with a review of the fundamentals: likelihoods, priors, and posteriors.
Next, we discuss Bayesian evidence, Bayes factors, odds, and model selection.
From there, we describe how posteriors are estimated using samplers such as Markov Chain Monte Carlo algorithms and nested sampling.
Finally, we generalize the formalism to discuss hyper-parameters and hierarchical models.
We include extensive appendices discussing the creation of credible intervals, Gaussian noise, explicit marginalization, posterior predictive distributions, and selection effects.
\end{abstract}

\end{frontmatter}

\section{Preface: why study Bayesian inference?}
Bayesian inference is an essential part of modern astronomy.
It finds particularly elegant application in the field of gravitational-wave astronomy thanks to the clear predictions of general relativity and the extraordinary simplicity with which compact binary systems are described.
An astrophysical black hole is completely characterized by just its mass and its dimensionless spin vector.
The gravitational waveform from a black hole binary is typically characterized by just fifteen parameters.
Since sources of gravitational waves are so simple, and since we have a complete theory describing how they emit gravitational waves, there is a direct link between data and model.
The significant interest in Bayesian inference within the gravitational-wave community reflects the great possibilities of this area of research.

Bayesian inference and parameter estimation are the tools that allow us to make statements about the Universe based on data.
In gravitational-wave astronomy, Bayesian inference is the tool that allows us to reconstruct sky maps of where a binary neutron star merged~\citep{GW170817}, to determine that GW170104 merged $\unit[880^{+450}_{-390}]{Mpc}$ away from Earth~\citep{gw170104}, and that the black holes in GW150914 had masses of $\unit[35^{+5}_{-3}]{M_\odot}$ and $\unit[33^{+3}_{-4}]{M_\odot}$~\citep{gw150914}.
We use it to determine the Hubble constant~\citep{Hubble}, to study the formation mechanism of black hole binaries~\citep{salvo,Stevenson,spin,GerosaBerti,FarrNature,Wysocki18,eccentricity}, and to probe how stars die~\citep{mass_uc,mass,r_and_p}.
Increasingly, Bayesian inference and parameter estimation are the language of gravitational-wave astronomy.
In this note, we endeavor to provide a primer on Bayesian inference with examples from gravitational-wave astronomy\footnote{This review focuses on Bayesian inference applied to audio-band gravitational waves from compact binary coalescence, the only source of gravitational waves yet detected. We note in passing that Bayesian inference has been applied to study gravitational waves from rotating neutron stars~\citep{Umstatter,Dupuis,KnownPulsars}, bursting sources~\citep{BayesWave,SMEE1,SMEE2}, and stochastic backgrounds~\citep{Mandic,Polarization,PolarizationSearch}.
Bayesian inference methods have also been developed for space-based observatories observing at millihertz frequencies~\citep{LISA1,LISA2} and for pulsar timing arrays operating at nanohertz frequencies~\citep{TempoNest,Tempo2}.}.

Before beginning, we highlight additional resources, useful for researchers interested in Bayesian inference in gravitational-wave astronomy.
\cite{SiviaSkilling} and \cite{Gregory} are useful references that are accessible to physicists and astronomers; see also the Springer Series in Astrostatistics~\citep{Springer1,Springer2,Springer3,Springer4}.
The chapter in~\cite{Springer2} by Loredo discusses hierarchical models, but refers to them as ``multilevel'' models~\citep{Loredo}.
Seasoned veterans may find \cite{gelman2013bayesian} to be a thorough reference.

This version of the paper includes updates, which corrects mistakes in the originally published version.
A summary of the changes are included in Appendix~\ref{erratum}.

\section{Fundamentals: likelihoods, priors, and posteriors}
A primary aim of modern Bayesian inference is to construct a posterior distribution
\begin{align}
p(\theta | d) .
\end{align}
Here, $\theta$ is the set of model parameters and $d$ is the data associated with a measurement\footnote{By referring to ``model parameters,'' we are implicitly acknowledging that we begin with some model.
Some authors make this explicit by writing the posterior as $p(\theta | d, M)$ where $M$ is the model.
(Other authors sometimes use $I$ to denote the model.)
We find this notation clunky and unnecessary since it goes without saying that one must always assume {\em some} model.
If/when we consider two {\em distinct} models, we add an additional variable to denote the model.}.
For illustrative purposes, let us say that $\theta$ are the 15 parameters describing a binary black hole coalescence and $d$ is the strain data from a network of gravitational-wave detectors.
The posterior distribution $p(\theta | d)$ is the probability density function for the continuous variable $\theta$ given the data $d$.
The probability that the true value of $\theta$ is between $(\theta', \theta'+d\theta')$ is given by $p(\theta'|d)d\theta'$.
It is normalized so that 
\begin{align}
\int d\theta \, p(\theta | d) = 1
\end{align}
The posterior distribution is what we use to construct credible intervals that tell us, for example, the component masses of a binary black hole event like GW150914.
For details about the construction of credible intervals, see Appendix~\ref{intervals}.

According to Bayes theorem, the posterior distribution is given by
\begin{align}\label{eq:posterior}
p(\theta | d) = \frac{
{\cal L}(d | \theta) \, 
\pi(\theta)
}
{\cal Z} .
\end{align}
Here, ${\cal L}(d|\theta)$ is the likelihood function of the data given the parameters $\theta$, $\pi(\theta)$ is the prior distribution for $\theta$, and ${\cal Z}$ is a normalization factor\footnote{
In this document we use different symbols for different distributions: $p$ for posteriors, ${\cal L}$ for likelihoods, and $\pi$ for priors.
We advocate this notation since it highlights what is what and makes formulas easy to read.
However, it is by no means standard, and some authors will use $p$ for any and all probability distributions.}\footnote{For now, we  treat the evidence as ``just'' a normalization factor, though, below we see that it plays an important role in model selection, and that it can be understood as a marginalized likelihood.} called the ``evidence''
\begin{align}\label{eq:evidence}
{\cal Z} \equiv \int d\theta {\cal L}(d | \theta) \, \pi(\theta) .
\end{align}

The likelihood function is something that we choose.
It is a description of the measurement.
By writing down a likelihood, we implicitly introduce a noise model.
For gravitational-wave astronomy, we typically assume a Gaussian-noise likelihood function that looks something like this
\begin{align}\label{eq:mu}
{\cal L}(d | \theta) = \frac{1}{2\pi\sigma^2} \exp\left(-\frac{1}{2}\frac{\left|d-\mu(\theta)\right|^2}{\sigma^2}\right) .
\end{align}
Here, $\mu(\theta)$ is a template for the gravitational strain waveform given $\theta$ and $\sigma$ is the detector noise.
Note that $\pi$ with no parentheses and no subscript is the mathematical constant, not a prior distribution.
There is no square root in the normalisation factor because $d$ is (typically) complex, which means that we are working with a two-dimensional Gaussian---the Whittle likelihood~\citep{Whittle}; see also~\cite{CornishRomano}.
This likelihood function reflects our assumption that the noise in gravitational-wave detectors is Gaussian\footnote{The Gaussian noise assumption is a good starting point for describing the strain noise in gravitational-wave detectors. The combined effect of many random noise processes tends to produce nearly Gaussian strain noise. Of course, the noise description can be generalized to include non-Gaussian glitches, drift over time, and instrumental lines all of which can be described by noise parameters; see, e.g.,~\citep{BayesLine,Rover}.}.
Note that the likelihood function is not normalized with respect to $\theta$ and so\footnote{Given that the likelihood is not normalized with respect to $\theta$, one might ask in what way it {\em is} normalized. The answer is that the likelihood is normalized with respect to the {\em data} $d$. Before we collect any data, the likelihood describes the chance of getting data $d$. It is a probability density function with units of inverse data. The integral over all possible $d$ is unity. Once we obtain actual data, $d$ is, of course, fixed.}
\begin{align}
\int d\theta \, {\cal L}(d | \theta) \neq 1 .
\end{align}
For a more detailed discussion of the Gaussian noise likelihood in the context of gravitational-wave astronomy, see Appendix~\ref{gaussian}.

Like the likelihood function, the prior is something we get to choose.
The prior incorporates our belief about $\theta$ before we carry out a measurement.
In some cases, there is an obvious choice of prior.
For example, if we are considering the sky location of a binary black hole merger, it is reasonable to choose an isotropic prior that weights each patch of sky as equally probable.
In other situations, the choice of prior is not obvious.
For example, before the first detection of gravitational waves, what would have been a suitable choice for the prior on the primary\footnote{The ``primary'' black hole is the heavier of two black holes in a binary, which is contrasted with the lighter ``secondary'' black hole.} black hole mass $\pi(m_1)$?
When we are ignorant about $\theta$, we often express our ignorance by choosing a distribution that is either uniform or log-uniform\footnote{A log uniform distribution is used when we do not know the order of magnitude of some quantity, for example, the energy density of primordial gravitational waves.}.

While $\theta$ may consist of a large number of parameters, we usually want to look at just one or two at a time.
For example, the posterior distribution for a binary black hole merger is a fifteen-dimensional\footnote{There are eight ``intrinsic'' parameters, which are fundamental properties of the binary: primary mass $m_1$, secondary mass $m_2$, primary dimensionless spin vector $\vec{s}_1$, and secondary dimensionless spin vector $\vec{s}_2$. The other seven parameters are ``extrinsic,'' relating to how we view the binary. The extrinsic parameters are: inclination angle $\iota$, polarization angle $\psi$, phase at coalescence $\phi_c$, right ascension $\text{RA}$, declination $\text{DEC}$, luminosity distance $D_L$, and time of coalescence $t$.} function that includes information about black hole masses, sky location, spins, etc.
What if we want to look at the posterior distribution for just the primary mass?
To answer this question we {\em marginalize} (integrate) over the parameters that we are not interested in (called ``nuisance parameters'') so as to obtain a marginalized posterior
\begin{align}\label{eq:marginalization}
p(\theta_i | d) = & \int \left(\prod_{k\neq i} d\theta_k\right) 
p(\theta | d) \\
= & \frac{
{\cal L}(d | \theta_i) \, \pi(\theta_i)
}
{\cal Z}
\end{align}
The quantity ${\cal L}(d | \theta_i)$ is called the ``marginalized likelihood.''
It can be expressed like so:
\begin{align}
{\cal L}(d | \theta_i) =
\int \left(\prod_{k\neq i} d\theta_k\right) \pi(\theta_k) 
\, {\cal L}(d | \theta)
\end{align}

When we marginalize over one variable $\theta_a$ in order to obtain a posterior on $\theta_b$, we are calculating our best guess for $\theta_b$ given uncertainty in $\theta_a$.
Speaking somewhat colloquially, if $\theta_a$ and $\theta_b$ are covariant, then marginalizing over $\theta_a$ ``injects'' uncertainty into the posterior for $\theta_b$.
When this happens, the marginalized posterior $p(\theta_b | d)$ is broader than the {\em conditional posterior} $p(\theta_b | d, \theta_a)$.
The conditional posterior $p(\theta_b | d, \theta_a)$ represents a slice through the $p(\theta_b | d)$ posterior at a fixed value of $\theta_a$.

\begin{figure}
\includegraphics[width=\linewidth]{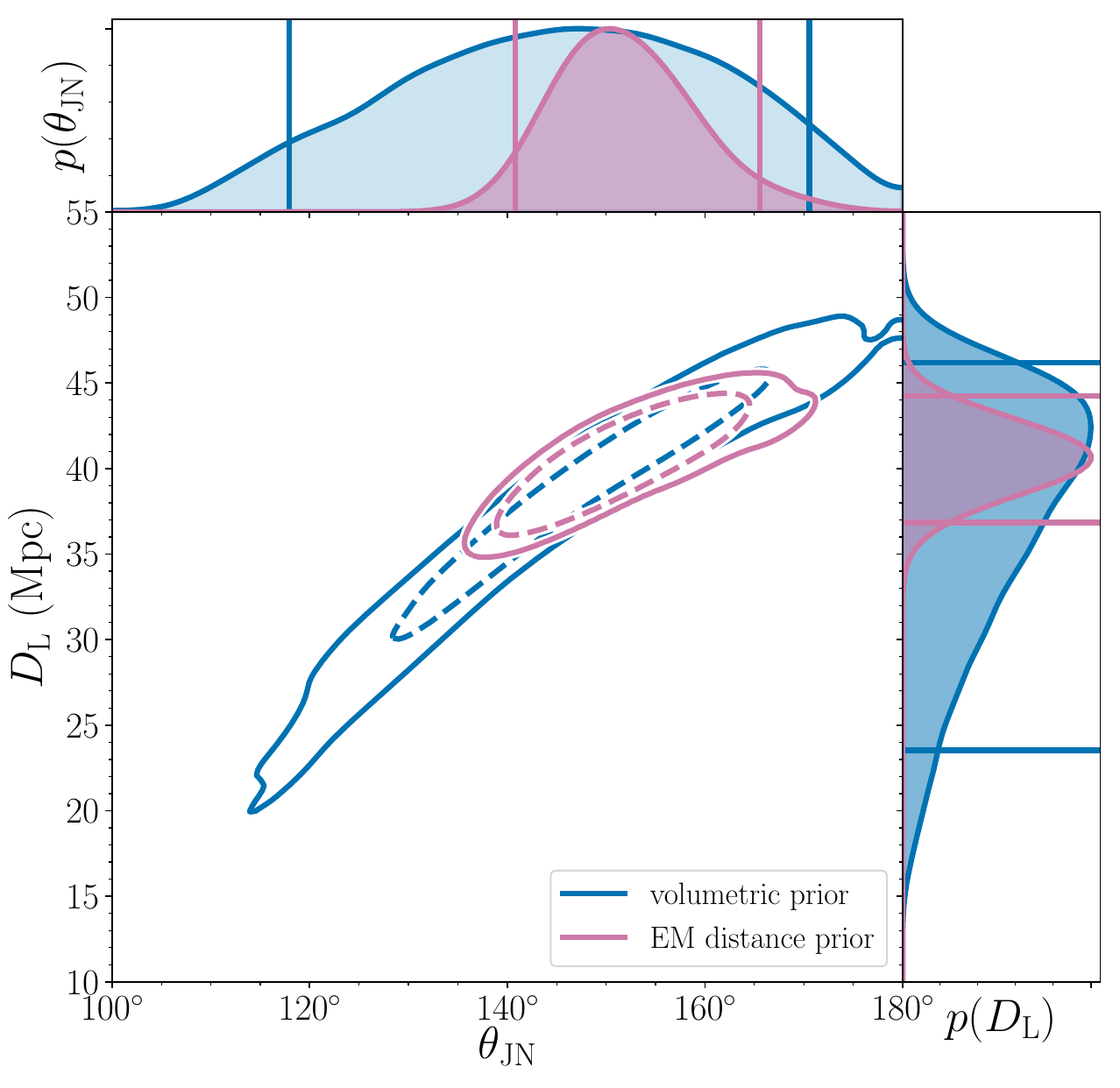}
\caption{The joint posterior for luminosity distance and inclination angle for GW170817 from~\citep{gw170817_properties}.
The blue contours show the credible region obtained using gravitational-wave data alone.
The purple contours show the smaller credible region obtained by employing a relatively narrow prior on distance obtained with electromagnetic measurements.
Publicly available posterior samples for this plot are available here:~\citep{P1800061}.
\label{fig:gw170817}
}
\end{figure}

This is nicely illustrated with an example.
There is a well-known covariance between the luminosity distance of a merging compact binary from Earth $D_L$ and the inclination angle $\theta_{JN}$.
For the binary neutron star coalescence GW170817, we are able to constrain the inclination angle much better when we use the known distance and sky location of the host galaxy compared to the constraint obtained using the gravitational-wave measurement alone\footnote{The viewing angle = $\Theta=\min(\theta_{JN},180^\circ-\theta_{JN}$ is constrained to be $<28^{\circ}$ with the electromagnetic counterpart, and $<55^{\circ}$ without it~\citep{GW170817}}.
Results from~\citep{gw170817_properties} are shown in Fig.~\ref{fig:gw170817}.

\section{Models, evidence and odds}
In Eq.~\ref{eq:evidence}, reproduced here, we defined the Bayesian evidence:
\begin{align}
{\cal Z} \equiv \int d\theta {\cal L}(d | \theta) \, \pi(\theta) \nonumber .
\end{align}
In practical terms, the evidence is a single number.
It usually does not mean anything by itself, but becomes useful when we compare one evidence with another evidence.
Formally, the evidence is a likelihood function.
Specifically, it is the completely marginalized likelihood function.
It is therefore sometimes denoted ${\cal L}(d)$ with no $\theta$ dependence.
However, we prefer to use ${\cal Z}$ to denote the fully marginalized likelihood function.

Above, we described how the evidence serves as a normalization constant for the posterior $p(\theta | d)$.
However, the evidence is also used to do model selection.
Model selection answers the question: which model is statistically preferred by the data and by how much?
There are different ways to think about models.
Let us return to the case of binary black holes.
We may compare a ``signal model'' in which we suppose that there is a binary black hole signal present in the data with a prior $\pi(\theta)$ to the ``noise model,'' in which we suppose that there is no binary black hole signal present.
While the signal model is described by the fifteen binary parameters $\theta$, the noise model is described by no parameters.
Thus, we can define a signal evidence ${\cal Z}_S$ and a noise evidence ${\cal Z}_N$
\begin{align}
{\cal Z}_S \equiv & \int d\theta {\cal L}(d | \theta) \, \pi(\theta) \\
{\cal Z}_N \equiv & {\cal L}(d | 0) ,
\end{align}
where
\begin{align}
{\cal L}(d | 0) \equiv \frac{1}{2\pi\sigma^2}
\exp\left(-\frac{1}{2}\frac{|d|^2}{\sigma^2}\right) .
\end{align}
The noise evidence ${\cal Z}_N$ is sometimes referred to as the ``null likelihood.''

The ratio of the evidence for two different models is called the Bayes factor.
In this example, the signal/noise Bayes factor is
\begin{align}
\text{BF}^S_N \equiv \frac{{\cal Z}_S}{{\cal Z}_N} .
\end{align}
It is often convenient to work with the log of the Bayes factor\footnote{A typical log evidence might be $-5000$, which evaluates to zero when exponentiated on a computer . Functions such as {\tt logsumexp} can be useful for combining evidences.}
\begin{align}
\log\text{BF}^S_N \equiv \log({\cal Z}_S) - \log({\cal Z}_N) .
\end{align}
When the absolute value of $\log\text{BF}$ is large, we say that one model is preferred over the other.
The sign of $\log\text{BF}$ tells us which model is preferred.
A threshold of $|\log\text{BF}|=8$ is often used as the level of ``strong evidence'' in favor of one hypothesis over another~\citep{Jeffreys61}.

The signal/noise Bayes factor is just one example of a Bayes factor comparing two models.
We can calculate a Bayes factor comparing identical models but with different priors.
For example, we can calculate the evidence for a binary black hole with a uniform prior on dimensionless spin and compare that to the evidence obtained using a zero-spin prior.
The Bayes factor comparing these models would tell us if the data prefer spin.
\begin{align}
{\cal Z}_{\text{spin}} = & \int d\theta {\cal L}(d | \theta) \, \pi(\theta) \\
{\cal Z}_{\text{no spin}} = & \int d\theta {\cal L}(d | \theta) \, \pi_\text{no spin}(\theta) .
\end{align}
Where $\pi_\text{no spin}(\theta)$ is a prior with zero spins.
The spin/no spin Bayes factor is 
\begin{align}
\text{BF}^{\text{spin}}_{\text{no spin}} = \frac{{\cal Z}_{\text{spin}}}{{\cal Z}_{\text{no spin}}} .
\end{align}

We may also compare two disparate signal models.
For example, we can compare the evidence for a binary black hole waveform predicted by general relativity (model $M_A$ with parameters $\theta$) with a binary black hole waveform predicted by some other theory (model $M_B$ with parameters $\nu$):
\begin{align}
{\cal Z}_A = & \int d\theta {\cal L}(d | \theta, M_A) \, \pi(\theta) \\
{\cal Z}_B = & \int d\nu {\cal L}(d | \nu, M_B) \, \pi(\nu) .
\end{align}
The $A/B$ Bayes factor is 
\begin{align}
\text{BF}^A_B = \frac{{\cal Z}_A}{{\cal Z}_B} .
\end{align}
Note that the number of parameters in $\nu$ can be different from the number of parameters in $\theta$.

Our presentation of model selection so far has been a bit fast and loose.
Formally, the correct metric to compare two models is not the Bayes factor, but rather the odds
\begin{align}
{\cal O}^A_B \equiv \frac{{\cal Z}_A}{{\cal Z}_B} \frac{\pi_A}{\pi_B} .
\end{align}
The odds is the product of the Bayes factor with the prior odds $\pi_A/\pi_B$, which describes our prior belief about the relative likelihood of hypotheses A and B.
In many practical applications, we set the prior odds to unity, and so the odds {\em is} the Bayes factor.
This practice is sensible in many applications where our intuition tells us: until we do this measurement both hypotheses are equally likely\footnote{There are some (fairly uncommon) examples where we might choose a different prior odds. 
For example, we may construct a model in which general relativity (GR) is wrong.
We may further suppose that there are multiple different ways in which it could be wrong, each corresponding to a different GR-is-wrong sub-hypothesis.
If we calculated the odds comparing one of these GR-is-wrong sub-hypotheses to the GR-is-right hypothesis, we would not assign equal prior odds to both hypotheses.
Rather, we would assign at most 50\% probability to the entire GR-is-wrong hypothesis, which would then have to be split among the various sub-hypotheses~\citep{Polarization}.}.

Bayesian evidence encodes two pieces of information.
First, the likelihood tells us how well our model fits the data.
Second, the act of marginalization tell us about the size of the volume of parameter space we used to carry out a fit.
This creates a sort of tension.
We want to get the best fit possible (high likelihood) but with a minimum prior volume.
A model with a decent fit and a small prior volume often yields a greater evidence than a model with an excellent fit and a huge prior volume.
In these cases, the Bayes factor penalizes the more complicated model for being too complicated.

This penalty is called an Occam factor.
It is a mathematical formulation of the statement that all else equal, a simple explanation is more likely than a complicated one.
If we compare two models where one model is a superset of the other---for example, we might compare general relativity and general relativity with non-tensor modes---and if the data are better explained by the simpler model, the log Bayes factor is typically modest, $\log\text{BF}\approx (-2, -1)$.
Thus, it is difficult to completely rule out extensions to existing theories.
We just obtain ever tighter constraints on the extended parameter space.

\section{Samplers}
Thanks to the creation of phenomenological gravitational waveforms (called ``approximants''), it is now computationally straightforward to make a prediction about what the data $d$ should look like given some parameters $\theta$.
That is a forward problem.
Calculating the posterior, the probability of parameters $\theta$ given the data as in Eq.~\ref{eq:posterior}, reproduced here, is a classic inverse problem\footnote{We note here a few early papers important in the development of Bayesian inference tools for gravitational-wave astronomy. Initial implementation of MCMC methods for spinning binaries was carried out in~\cite{VanDerSluys1}. The first demonstration of Bayesian parameter estimation for spinning binaries was performed in~\cite{VanDerSluys2}. \cite{VeitchVecchio}, demonstrated Bayesian model selection for compact binaries.}
\begin{align}
p(\theta | d) = \frac{
{\cal L}(d | \theta) \, 
\pi(\theta)
}
{\cal Z} . \nonumber
\end{align}
In general, inverse problems are computationally challenging compared to forward problems.
To illustrate why let us imagine that we wish to calculate the posterior probability for the fifteen parameters describing a binary black hole merger.
If we do this naively, we might create a grid with ten bins in every dimension and evaluate the likelihood at each grid point.
Even with this coarse resolution, our calculation suffers from ``the curse of dimensionality.'' 
It is computationally prohibitive to carry out $10^{15}$ likelihood evaluations.
The problem becomes worse as we add dimensions.
As a rule of thumb, brute-force bin approaches become painful once one exceeds three dimensions.

The solution is to use a stochastic sampler, (although recent work has shown progress carrying out these calculations using the alternative technique of iterative fitting~\citep{RapidPE1,RapidPE2}).
Commonly used sampling algorithms can be split into two broad categories of method: Markov-chain Monte Carlo (MCMC)~\citep{Metropolis1953,Hastings1970} and nested sampling~\citep{Skilling2004}.
These algorithms generate a list of posterior samples $\{\theta\}$ drawn from the posterior distribution such that the number of samples on the interval $(\theta, \theta + \Delta\theta) \propto p(\theta)$~\citep{lalinference}.
Some samplers also produce an estimate of the evidence.
We can visualize the posterior samples as a spreadsheet.
Each column is a different parameter, for example, primary black hole mass, secondary black hole mass, etc.
For binary black hole mergers, there are typically fifteen columns.
Each row represents a different posterior sample.

Posterior samples have two useful properties.
First, they can be used to compute expectation values of quantities of interest since~\citep{Hogg2018}
\begin{equation}
\langle f(x) \rangle_{p(x)} = 
\int dx \, p(x) \, f(x) \approx \frac{1}{n_s}
\sum_k^{n_s} f(x_k) .
\label{eq:sampling}
\end{equation}
Here $p(x)$ is the posterior distribution that we are sampling, $f(x)$ is some function we want to find the expectation value of, and the sum over $k$ runs over $n_{s}$ posterior samples.
Below, Eq.~\ref{eq:sampling} will prove useful simplifying our calculation of the likelihood of data given hyper-parameters.

The second useful property of posterior samples is that, once we have samples from an N-dimensional space, we can generate the marginalized probability for any subset of the parameters by simply selecting the corresponding columns in our spreadsheet.
This property is used to help visualize the output of these samplers by constructing ``corner plots,'' which show the marginalized one- and two-dimensional posterior probability distributions for each of the parameters.
For an example of a corner plot, see Fig.~\ref{fig:gw170817}.
A handy python package exists for making corner plots~\citep{corner}.

\subsection{MCMC}
Markov chain Monte Carlo sampling was first introduced in~\cite{Metropolis1953} and extended in~\cite{Hastings1970}.
For a recent overview of MCMC methods in astronomy, see \cite{Sharma}.
In MCMC methods, particles undergo a random walk through the posterior distribution where the probability of moving to any given point is determined by the transition probability of the Markov chain.
By noting the position of the particles---or ``walkers'' as they are sometimes called---at each iteration, we generate draws from the posterior probability distribution.

There are some subtleties that must be considered when using MCMC samplers.
First, the early-time behavior of MCMC walkers is strongly dependent on the initial conditions.
It is therefore necessary to include a ``burn-in'' phase to ensure that the walker has settled into a steady state before beginning to accumulate samples from the posterior distribution.
Once the walker has reached a steady state, the algorithm can continue indefinitely and so it is necessary for the user to define a termination condition.
This is typically chosen to be when enough samples have been acquired for the user to believe an accurate representation of the posterior has been obtained.
Thus, MCMC requires a degree of artistry, developed from experience.

Additionally, the positions of a walker in a chain are often autocorrelated.
Because of this correlation, the positions of the walkers do not represent a faithful sampling from the posterior distribution.
If no remedy is applied, the width of the posterior distribution is underestimated.
It is thus necessary to ``thin'' the chain by selecting samples separated by the autocorrelation length of the chain.

Markov chain Monte Carlo walkers can also fail to find multiple modes of a posterior distribution if there are regions of low posterior probability between the modes.
However, this can be mitigated by running many walkers which begin exploring the space at different points.
This also demonstrates a simple way to parallelize MCMC computations to quickly generate many samples.
Many variants of MCMC sampling have been proposed in order to improve the performance of MCMC algorithms with respect to these and other issues.
For a more in-depth discussion of MCMC methods see, e.g., chapter 11 of~\citep{gelman2013bayesian}, or~\citep{Hogg2018}.
The most widely used MCMC code in astronomy is {\sc emcee}~\citep{emcee}\footnote{\href{http://dfm.io/emcee/}{http://dfm.io/emcee/}}.

\subsection{Nested sampling}
The first widely used alternative to MCMC, was introduced by~\citeauthor{Skilling2004} in 2004.
While MCMC methods are designed to draw samples from the posterior distribution, nested sampling is designed to calculate the evidence.
Generating samples from the posterior distribution is a by-product of the nested sampling evidence calculation algorithm.
By weighting each of the samples used to calculate the evidence by the posterior probability of the sample, nested samples are converted into posterior samples.

Nested sampling works by populating the parameter space with a set of ``live points'' drawn from the prior distribution.
At each iteration, the lowest likelihood point is removed from the set of live points and new samples are drawn from the prior distribution until a point with higher likelihood than the removed point is found.
The evidence is evaluated by assigning each removed point a prior volume and then computing the sum of the likelihood multiplied by the prior volume for each sample.

Since the nested sampling algorithm continually moves to higher likelihood regions, it is possible to estimate an upper limit on the evidence at each iteration.
This is done by imagining that the entire remaining prior volume has a likelihood equal to that of the highest likelihood live point.
This is used to inform the termination condition for the nested sampling algorithm.
The algorithm stops when the current estimate of the evidence is above a certain fraction of the estimated upper limit\footnote{In practice this is expressed as the difference between the calculated log evidence and the upper limit of the log evidence.}.
Unlike MCMC algorithms nested sampling is not straightforwardly parallelizable, and posterior samples do not accumulate linearly with run time.

\section{Hyper-parameters and hierarchical models}
As more and more gravitational-wave events are detected, it is increasingly interesting to study the {\em population properties} of binary black holes and binary neutron stars.
These are the properties common to all of the events in some set.
Examples include the neutron star equation of state and the distribution of black hole masses.
Hierarchical Bayesian inference is a formalism, which allows us to go beyond individual events in order to study population properties\footnote{Possibly the earliest papers proposing to measure {\em distributions} of gravitational-wave parameters are~\citep{0912.1074,0912.5531} while hierarchical Bayesian inference was introduced to study the population properties of sources of gravitational waves in~\citep{1209.6286}.}.

The population properties of some set of events is described by the shape of the prior.
For example, two population synthesis models might yield two different predictions for the prior distribution of the primary black hole mass $\pi(m_1)$.
In order to probe the population properties of an ensemble of events, we make the prior for $\theta$ conditional on a set of ``hyper-parameters'' $\Lambda$
\begin{align}
\pi(\theta | \Lambda) .
\end{align}
The hyper-parameters parameterize the shape of the prior distribution for the parameters $\theta$.
An example of a (parameter, hyper-parameter) relationship is ($\theta$ = primary black hole mass $m_1$, $\Lambda$ = the spectral index of the primary mass spectrum $\alpha$).
In this example
\begin{align}
\pi(m_1 | \alpha) \propto m_1^\alpha .
\end{align}

A key goal of population inference is to estimate the posterior distribution for the hyper-parameters $\Lambda$.
In order to do this, we marginalize over the entire parameter space $\theta$ in order to obtain a marginalized likelihood.
\begin{align}
{\cal L}(d | \Lambda) = \int d\theta \,
{\cal L}(d | \theta) \, \pi(\theta | \Lambda) .
\end{align}
Normally, we would call this completely marginalized likelihood an evidence, but because it still depends on $\Lambda$, we call it the likelihood for the data $d$ given the hyper-parameters $\Lambda$.
The hyper-posterior is given simply by
\begin{align}
p(\Lambda | d) = \frac{
{\cal L}(d | \Lambda) \, 
\pi(\Lambda)
}
{\int d\Lambda \, {\cal L}(d | \Lambda) \, 
\pi(\Lambda)} .
\end{align}
Note that we have introduced a hyper-prior $\pi(\Lambda)$, which describes our prior belief about the hyper-parameters $\Lambda$.
The term in the denominator
\begin{align}
{\cal Z}_\Lambda \equiv \int d\Lambda \, {\cal L}(d | \Lambda) \, 
\pi(\Lambda)
\end{align}
is the ``hyper-evidence,'' which we denote ${\cal Z}_\Lambda$ in order to distinguish it from the regular evidence ${\cal Z}_\theta$.
In Appendix~\ref{ppd} we discuss posterior predictive distributions (PPD), which represent the updated prior on $\theta$ in light of the data $d$ and given some hyper-parameterization.

We now generalize the discussion of hyper-parameters in order to handle the case of $N$ independent events.
In this case, the total likelihood for all $N$ events ${\cal L}_\text{tot}$ is simply the product of each individual likelihood
\begin{align}\label{eq:product}
{\cal L}_\text{tot}(\vec{d} | \vec\theta) = \prod_i^N {\cal L}(d_i | \theta_i) .
\end{align}
Here, we use vector notation so that $\vec{d}$ is the set of measurements of $N$ events, each of which has its own parameters, which make up the vector $\vec\theta$.
Since we suppose that every event is drawn from the same population prior distribution---hyper-parameterized by $\Lambda$---the total marginalized likelihood is
\begin{align}\label{eq:Ltot}
{\cal L}_\text{tot}(\vec{d} | \Lambda) = \prod_i^N \int d\theta_i \, 
{\cal L}(d_i | \theta_i) 
\, \pi(\theta_i | \Lambda) .
\end{align}
The associated (hyper-) posterior is
\begin{align}\label{eq:hyperp}
p_\text{tot}(\Lambda | \vec{d}) = \frac{
{\cal L}_\text{tot}(\vec{d} | \Lambda) \, \pi(\Lambda)
}
{\int d\Lambda \, {\cal L}_\text{tot}(\vec{d} | \Lambda) \, \pi(\Lambda)} .
\end{align}
The denominator, of course, is the total hyper-evidence.
\begin{align}\label{eq:Ztot}
{\cal Z}_\Lambda^\text{tot} = \int d\Lambda \, {\cal L}_\text{tot}(\vec{d} | \Lambda) \, \pi(\Lambda)
\end{align}
We may calculate the Bayes factor comparing different hyper-models in the same way that we calculate the Bayes factor for different models.

Examining Eq.~\ref{eq:Ztot}, we see that the total hyper-evidence involves a large number of integrals.
For the case of binary black hole mergers, every event has $15$ parameters, and so the dimension of the integral is $15N+M$ taking where $M$ is the number of hyper-parameters in $\Lambda$.
As $N$ gets large, it becomes difficult to sample such a large prior volume all at once.
Fortunately, it is possible to break the integral into individual integrals for each event, which are then combined through a process sometimes referred to as ``recycling.''

It turns out that the total marginalized likelihood in Eq.~\ref{eq:Ltot} can be written like so 
\begin{align}\label{eq:boxed}
\boxed{
{\cal L}_\text{tot}(\vec{d} | \Lambda) = \prod_i^N
\frac{{\cal Z}_{\o}(d_i)}{n_i}
\sum_k^{n_i} \frac{\pi(\theta^k_i|\Lambda)}{\pi(\theta^k_i|{\o})} 
}.
\end{align}
Here, the sum over $k$ is a sum over the $n_i$ posterior samples associated with event $i$.
The posterior samples for each event are generated with some default prior $\pi(\theta_k|\o)$.
The default prior is ultimately canceled from the final answer, so it not so important what we choose for the default prior so long as it is sufficiently uninformative.
Using the $\o$ prior, we obtain an evidence ${\cal Z}_{\o}$.
In this way, we are able to analyze each event individually before recycling the posterior samples to obtain a likelihood of the data given $\Lambda$.

To see where this formula comes from, we note that 
\begin{align}
p(\theta_i | d_i, \o) = 
\frac{
{\cal L}(d_i | \theta_i) \, \pi(\theta_i | \o)
}
{{\cal Z}_{\o}(d_i)}
\end{align}
Rearranging terms,
\begin{align}
{\cal L}(d_i | \theta_i)  = 
 {\cal Z}_{\o}(d_i) \, \frac{p(\theta_i | d_i, \o)
}
{\pi(\theta_i | \o)} .
\end{align}
Plugging this into Eq.~\ref{eq:Ltot}, we obtain\footnote{One ``recycles'' the posterior samples generated using the the $\pi(\theta_i|\o)$ prior in order to do something new with the hyper-parameterized prior $\pi(\theta_i|\Lambda)$.}
\begin{align}
{\cal L}_\text{tot}(\vec{d} | \Lambda) = \prod_i^N \int d\theta_i \, 
p(\theta_i | d_i, \o) \, {\cal Z}_{\o}(d_i) \,
\frac{\pi(\theta_i | \Lambda)}{\pi(\theta_i | \o)} .
\end{align}
Finally, we use Eq.~\ref{eq:sampling} to convert the integral over $\theta_i$ to a sum over posterior samples, thereby arriving at Eq.~\ref{eq:boxed}.

All of the results derived up until this point ignore selection effects where an event with parameters $\theta_1$ is easier to detect than an event with parameters $\theta_2$.
There are cases where selection effects are important.
For example, the visible volume for binary black hole mergers scales as approximately $V\propto M^{2.1}$, which means that higher mass mergers are relatively easier to detect than lower mass mergers~\citep{mass_uc}.
In Appendix~\ref{selection}, we show how this method is extended to accommodate selection effects.

\section{acknowledgments}
This document is the companion paper to a lecture at the 2018 OzGrav Inference Workshop held July 16-18, 2018 at Monash University in Clatyon, Australia.
Thank you to the organizers: Greg Ashton, Paul Lasky, Hannah Middleton, and Rory Smith.
This workshop was supported by OzGrav through the Australian Research Council CE170100004.
For helpful comments on a draft of this manuscript, we thank Sylvia Biscoveanu, Paul Lasky, Nikhil Sarin, and Shanika Galaudage.
We are indebted to Will Farr who clarified our understanding of selection effects and to Patricia Schmidt who helped our understanding of phase marginalization.
We thank Rory Smith and John Veitch for drawing our attention to the explicit distance marginalization work of Leo Singer.
Finally, we thank the anonymous referee for helpful suggestions, which improved the manuscript.
ET and CT are supported by CE170100004.
ET is supported by FT150100281.

\begin{appendix}
\section{Credible intervals}\label{intervals}
It is often convenient to use the posterior to construct ``credible intervals,'' regions of parameter space containing some fraction of posterior probability.
(Note that Bayesian inference yields credible intervals while frequentist inference yields {\em confidence intervals}.)
For example, one can plot one-, two-, and three-sigma contours.
By definition, a two-sigma credible region includes 95\% of the posterior probability, but this requirement does not uniquely determine a single credible region.
One well-motivated method for constructing confidence intervals is the highest posterior density interval (HPDI) method.

We can visualize the HPDI method as follows.
We draw a horizontal line through a posterior distribution and calculate the area of above the line.
If we move the line down, the area goes up.
If we place the line such that the area is 95\%, then the posterior above the line is the HPDI two-sigma credible interval.
In general, the HPDI is neither symmetric nor unimodal.
The advantage of HPDI over other methods is that it yields the minimum width credible interval.
This method is sometimes referred to as ``draining the bathtub.''

Another commonly used method for calculating credible intervals is to construct symmetric intervals.
Symmetric credible intervals are constructed using the cumulative distribution function,
\begin{equation}
P(x) = \int_{-\infty}^{x} dx' \, p(x').
\end{equation}
The $X\%$ credible region is the region
\begin{equation}
\frac{1}{2} \left(1 - \frac{X}{100}\right) < P(x) < \frac{1}{2} \left(1 + \frac{X}{100}\right).
\end{equation}
While symmetric credible intervals are simpler to construct than HPDI, particularly from samples drawn from a distribution, they can be misleading for multi-modal distributions and for distributions which peak near prior boundaries.

Credible intervals are useful for testing and debugging inference projects.
Before applying an inference calculation to real data, it is useful to test it on simulated data.
The standard test, see, e.g.,~\cite{Sidery}, is to simulate data $d$ according to parameters $\theta_\text{true}$ drawn at random from the prior distribution $\pi(\theta)$.
Then, we analyze this data in order to obtain a posterior $p(\theta | d)$.
The true value should fall inside the 90\% credible interval 90\% of the time.
Testing that this is true provides a powerful validation of the inference algorithm.
Note that we do not expect the posterior to peak precisely at $\theta_\text{true}$, just within the one-sigma region.

\section{Gaussian noise likelihood}\label{gaussian}
In this appendix, we introduce additional notation that is helpful for talking about the Gaussian noise likelihood frequently used in gravitational-wave astronomy.
In the main body of the manuscript, $d$ has been taken to represent data.
Now, we take $d$ to represent the Fourier transform of the strain time series $d(t)$ measured by a gravitational-wave detector.
In the language of computer programming,
\begin{align}
d = {\tt fft}\left(d(t)\right) / f_s ,
\end{align}
where $f_s$ is the sampling frequency and ${\tt fft}$ is a Fast Fourier transform.
The noise in each frequency bin is characterized by the single-sided noise power spectral density $P(f)$, which is proportional to strain squared and which has units of $\unit[]{Hz^{-1}}$.

The likelihood for the data in a single frequency bin $j$ given $\theta$ is
\begin{align}
{\cal L}(d_j | \theta) = \frac{1}{2\pi P_j} \exp\left(-2\Delta f \frac{\left|d_j-\mu_j(\theta)\right|^2}{P_j}\right) .
\end{align}
Here $\Delta f$ is the frequency resolution.
The factor of $2 \Delta f$ comes about from a factor of $1/2$ in the normal distribution and a factor of $4 \Delta f$ needed to convert the square of the Fourier transforms into units of one-sided power spectral density.
Note that the normalisation factor does not contain a square root because the data are complex, and so the Gaussian is a two-dimensional Whittle likelihood~\citep{Whittle}; see also~\cite{CornishRomano}.
The template $\mu(\theta)$ is related to the metric perturbation $h_{+,\times}(\theta)$ via antenna response factors $F_{+,\times}$~\citep{Warren}
\begin{align}
    \mu(\theta) = 
    F_+(\text{RA},\text{DEC},\psi) h_+(\theta)
    + F_\times(\text{RA},\text{DEC},\psi) h_\times(\theta)
\end{align}

Gravitational-wave signals are typically spread over many ($M$) frequency bins.
Assuming the noise in each bin is independent,  the combined likelihood is a product of the likelihoods for each bin
\begin{align}
{\cal L}({\bf d} | \theta) = & \prod_j^M {\cal L}(d_j | \theta)
\label{eq:frequency_product}
\end{align}
Here ${\bf d}$ is the set of data including all frequency bins and $d_j$ represents the data associated with frequency bin $j$.
If we consider a measurement with multiple detectors, the product over $j$ frequency bins gains an additional index $l$ for each detector.
Combining data from different detectors is like combining data from different frequency bins.

It is frequently useful to work with the log likelihood, which allows us to replace products with sums of logs.
The log also helps dealing with small numbers.
The log likelihood is
\begin{align}
\log{\cal L}({\bf d} | \theta) = & \sum_j^M \log{\cal L}(d_j | \theta) \nonumber\\
= & -\sum_j \log\left(2\pi P_j\right) 
-2 \Delta f \sum_j \frac{\left|d-\mu(\theta)\right|^2}{P_j} \nonumber\\
= & \Psi 
-\frac{1}{2} \langle d-\mu(\theta), d-\mu(\theta) \rangle \nonumber .
\end{align}
In the last line, we define the noise-weighted inner product\footnote{Following the convention of gravitational-wave astronomy, our inner product is real by construction. However, below it will be useful to define a complex-valued inner product; see Eq.~\ref{eq:inner_product_complex}.}~\citep{CutlerFlanagan}
\begin{align}\label{eq:inner_product}
\langle a,b \rangle \equiv 4\Delta f
\sum_j \Re\left( \frac{a_j^* b_j}{P_j} \right) ,
\end{align}
and the constant
\begin{align}
\Psi \equiv -\sum_j \log\left(2\pi P_j\right) .
\end{align}
Since constants do not change the shape of the log likelihood we often ``leave off'' this normalizing term and work with log likelihood minus $\Psi$.
This is permissible as long as we do so consistently because when we take the ratio of two evidences---or equivalently, the difference of two log evidences---the $\Psi$ factor cancels anyway.
For the remainder of this appendix, we set $\Psi=0$.
Now that we have introduced the inner product notation, we are going to stop bold-facing the data $d$ as it is implied that we are dealing with many frequency bins.

Using the inner product notation, we may expand out the log likelihood
\begin{align}\label{eq:GaussianNoiseLikelihood}
\log {\cal L}(d|\theta) = & 
-\frac{1}{2} \left[
\langle d,d \rangle 
-2\langle d,\mu(\theta) \rangle
+ \langle \mu(\theta),\mu(\theta) \rangle
\right] \nonumber\\
= & 
-\frac{1}{2} \left[
-2\log{\cal Z}_N
-2\ks(\theta)
+ \rho_\text{opt}^2(\theta)
\right] \nonumber\\
= & 
\log{\cal Z}_N
+ \ks(\theta)
- \frac{1}{2} \rho_\text{opt}^2(\theta) .
\end{align}
We see that the log likelihood can be expressed with three terms.
The first is proportional to the log noise evidence
\begin{align}
-2\log{\cal Z}_N \equiv \langle d,d \rangle .
\end{align}
For debugging purposes, it is useful to keep in mind that if we calculate $-\log{\cal Z}_N$ on actual Gaussian noise (with $\Psi=0$), we expect a typical value nearly equal to the number of frequency bins $M$ (multiplied by the number of detectors) since each term in the inner product contributes a value close to unity\footnote{Specifically, the distribution of an ensemble of independent $-\ln {\cal Z}_N$ is a normal distribution with mean $M$ and width $M^{1/2}$ where $M$ is the number of frequency bins (multiplied by the number of detectors). This follows from the central limit theorem.}.
We skip over the second term $\ks$ for a moment.
The third term is the optimal matched filter signal-to-noise ratio squared
\begin{align}
\rho_\text{opt}^2 \equiv \langle \mu,\mu \rangle .
\end{align}
Returning now to the second term, we express $\ks$ as the product of the matched filter signal-to-noise ratio and the optimal signal-to-noise ratio
\begin{align}\label{eq:x}
\ks \equiv & \langle d,\mu \rangle \nonumber\\
= & \rho_\text{mf} \, \rho_\text{opt} ,
\end{align}
where
\begin{align}
    \rho_\text{mf} \equiv 
    \frac{\langle d, \mu\rangle}{\langle\mu,\mu\rangle^{1/2}} .
\end{align}

Readers familiar with gravitational-wave astronomy are likely acquainted with the concept of matched filtering, which is the maximum likelihood technique for gravitational-wave detection.
By writing the likelihood in this way, we highlight how parameter estimation is related to matched filtering.
Rapid evaluation of the likelihood function in Eq.~\ref{eq:GaussianNoiseLikelihood} has been made possible through reduced order methods~\citep{smith,purrer,canizares}.

\section{Explicitly Marginalized Likelihoods}
The most computationally expensive step in computing the likelihood for compact binary coalescences is creating the waveform template ($\mu$ in Eq.~\ref{eq:mu}).
This is done in two steps.
The first step is to use the {\em intrinsic parameters} to calculate the metric perturbation.
The second (much faster) step is to use the {\em extrinsic parameters} to project the metric perturbation onto the detector response tensor.
In some cases, it is possible to reduce the dimensionality of the inverse problem---thereby speeding up calculations and improving convergence---by using a likelihood, which explicitly marginalizes over extrinsic parameters.
The improvement is especially marked for comparatively weak signals, which can be important for population studies; see, e.g.,~\citep{tbs}.
In this appendix, we show how to calculate ${\cal L}_\text{marge}$---a likelihood, which explicitly marginalize over coalescence time, phase at coalescence, and/or luminosity distance.
We continue with notation introduced in Appendix~\ref{gaussian}.

\subsection{Time marginalization}
In this subsection, we follow~\cite{T1400460} to derive a likelihood, which explicitly marginalizes over time of coalescence $t$.
Given a waveform with a reference coalescence time of $t_0$, we can calculate the waveform at some new coalescence time $t$ by multiplying by the appropriate phasor:
\begin{equation}
\begin{split}
	\mu_{j}(t)
    &= \mu_{j}(t_0) \exp\left(-2 \pi i j \frac{(t - t_0)}{T}\right) .
\end{split}
\label{eq:TimeDependence}
\end{equation}
Here $T=1/\Delta f$ is the duration of data segment and $j$ is the index of the frequency bin as in Appendix~\ref{gaussian}.
It is understood that $\mu$ is a function of whatever parameters we are not explicitly marginalizing over.
We can therefore write $\ks$ (see Eq:~\ref{eq:x}) as 
\begin{equation}
\begin{split}
	\ks(t) \equiv & \langle d, \mu(t) \rangle \\
    = & 4\Delta f
\Re \sum_{j}^{M} \frac{d_{j}^{*} \mu_j( t_0)}{P_{j}} \exp\left(-2 \pi i j  \frac{(t - t_0)}{T}\right).
\end{split}
\end{equation}
However this sum is the discrete Fourier transform.
By recasting this equation in terms of the fast Fourier transform ${\tt fft}$, it is possible to take advantage of a highly optimized tool. 

We discretize $t-t_0 = k \Delta t$ where $k$ takes on integer values between 0 and $M = T / \Delta t$.
Having made this definition, marginalizing over coalescence time becomes summing over $k$.
The variable $\ks$ is a function of (discretized) coalescence time $k$.
We can write in terms of a fast Fourier transform.
\begin{equation}
\begin{split}
	\ks(k) &= 4\Delta f
\Re \sum_{j}^{M} \frac{d_{j}^{*} \mu_j( t_0)}{P_{j}} \exp\left(-2 \pi i j \frac{k}{M}\right) \\
	&= 4 \Delta f \Re \, {\tt fft}_{k}\left(\frac{d_{j}^{*} \mu_j(t_0)}{P_{j}} \right).
\end{split}
\end{equation}
Here ${\tt fft}_k$ refers to the $k$ bin of a fast Fourier transform.

The other terms in~\ref{eq:GaussianNoiseLikelihood} are independent of the time at coalescence of the template.
The marginalized likelihood is therefore
\begin{equation}
\begin{split}
	\log &\mathcal{L}_\text{marg}^t = \log \int_{t_0}^{t_0 + T} dt \, \mathcal{L}(\theta, t) \pi(t) \\
    &= \log{\cal Z}_N - \frac{1}{2} \rho_\text{opt}^2(\theta) + \log \int_{t_0}^{t_0 + T} dt \, e^{\ks(\theta, t)} \pi(t) \\
    &= \log{\cal Z}_N - \frac{1}{2} \rho_\text{opt}^2(\theta) + \log \sum_{k}^{M} e^{\ks(\theta, k)} \pi_k ,
\end{split}
\end{equation}
where $\pi_k=\pi(t) \Delta t$ is the prior on the discretized coalescence time.

Caution should be taken to avoid edge effects.
If we employ a naive prior, the waveform will exhibit unphysical wrap-around.
Similarly, care must be taken to ensure that the time-shifted waveform is consistent with time-domain data conditioning, e.g., windowing.
(This is usually not a problem for confident detections because the coalescence time is well-known and so the segment edges can be avoided.)
A good solution is to choose a suitable prior, which is uniform over some values of $k$, but with some values set to zero in order to prevent the signal from wrapping around the edge of the data segment.
Note that Eq.~\ref{eq:TimeDependence} breaks down for when the detector changes significantly over $T$ due to the rotation of the Earth.
It can also fail in the high signal-to-noise ratio limit when the $t$ array becomes insufficiently fine-grained.

\subsection{Phase marginalization}
In this subsection, we follow~\cite{T1300326} (see also~\cite{lalinference}) to derive a likelihood, which explicitly marginalizes over phase of coalescence $\phi_c$.
To begin, we assume a gravitational waveform approximant consisting entirely of the dominant $\ell=2,|m|=2$ modes so that\footnote{
The variables $\mu_{22}$ and $\mu_{2-2}$ are defined like so
\begin{align}
    \mu_{\ell m} \equiv & F_+ \Re\Big( h_{\ell m}(\theta)
    \, _{-2} Y_{\ell m}(\iota, \phi)\Big) \nonumber\\
    + & F_\times \Im\Big( h_{\ell m}(\theta) 
    \, _{-2}Y_{\ell m}(\iota, \phi)\Big) .
\end{align}
They depend on the metric perturbation $h_{\ell m}$ and the antenna response functions $F_{+,\times}$.
The variable $_{-2}Y_{\ell m}(\iota,\phi)$ is a spin-weighted spherical harmonic function, evaluated the inclination angle $\iota$ and aziumuthal angle $\phi$ of the observer.
Without loss of generality, we can set $\phi=0$, which establishes a coordinate frame.
Having defined this frame, we may rotate the binary by the phase of coalescence $\phi_c$ in order to change the phase of the signal observed at Earth.}
\begin{align}
    \mu = \mu_{22}+\mu_{2-2} ,
\end{align}
This is a valid assumption, e.g., for the widely used waveform approximants---e.g., {\sc TaylorF2}~\citep{TaylorF2}, {\sc IMRPhenomD}~\citep{IMRPhenomD}, {\sc IMRPhenomP}~\citep{IMRPhenomP}---but not for waveforms that employ higher order modes, e.g.,~\citep{Blackman}.
Given this approximation\footnote{We emphasize that the phase at coalescence is distinct from $\phi$, the azimuthal angle to the observer in the source frame, which transforms differently
\begin{align}
    \mu(\phi) = e^{2i\phi} \mu_{22}(\phi=0) + 
    e^{-2i\phi} \mu_{2-2}(\phi=0) .
\end{align}
The variable $\phi_c$ calibrates the time evolution of the gravitational waveform observed at Earth, while $\phi$ describes how the the waveform  varies at a fixed time for observers at different spatial locations (corresponding to different azimuthal angles). 
},
\begin{align}\label{eq:phi}
    \mu(\phi_c) = e^{2i\phi_c} \mu(\phi_c=0) .
\end{align}

The optimal signal-to-noise ratio $\rho_\text{opt}$ is invariant under rotations in $\phi_c$.
However the matched filter signal-to-noise ratio is not.
Thus, the phase-marginalized  likelihood is
\begin{equation}
\begin{split}
  &\mathcal{L}_\text{marg}^{\phi_c} = {\cal Z}_N - \exp\left(\frac{1}{2} \rho_\text{opt}^2\right) \\
  &+ \int_{0}^{2\pi} d\phi_c \, \exp\Big(\frac{1}{2}
  \big\langle d, \mu(\phi_c)\big\rangle + \frac{1}{2}\big\langle \mu(\phi_c) , d \big\rangle \Big) \pi(\phi_c) .
\end{split}
\end{equation}
Using Eq.~\ref{eq:phi}, we can rewrite the phase-marginalized likelihood
\begin{equation}
\begin{split}
  \mathcal{L}_\text{marg}^{\phi_c} = \int_{0}^{2\pi} d\phi_c \, & \exp\Big(\frac{1}{2}
  \big\langle d, \mu(\phi_c=0)\big\rangle_\mathbb{C}\exp(2i\phi_c) + \nonumber\\
  & \frac{1}{2}\big\langle \mu(\phi_c=0) , d \big\rangle_\mathbb{C}\exp(-2i\phi_c) \Big) \pi(\phi_c) \nonumber\\
  & + ...
\end{split}
\end{equation}
The parts that do not depend on $\phi_c$ are implied by the ellipsis.
Here we introduce the ``complex inner product'' denoted with a subscript $\mathbb{C}$.
\begin{align}\label{eq:inner_product_complex}
\langle a,b \rangle_\mathbb{C} \equiv 4\Delta f
\sum_j \left( \frac{a_j^* b_j}{P_j} \right) ,
\end{align}
which is identical to the regular inner product defined in Eq.~\ref{eq:inner_product} except we do not take the real part in order to preserve phase information that will be useful later on.
Employing a uniform prior on $\phi_c$ and grouping terms, the integral can be rewritten yet again
\begin{equation}
  \mathcal{L}_\text{marg}^{\phi_c} = \int_{0}^{2\pi} \frac{d\phi_c}{2\pi} \, \exp\Big(A\cos(2\phi_c) + B\sin(2\phi_c) \Big) + ...
\end{equation}
where
\begin{align}
    A \equiv & \Re \big\langle d, \mu(\phi_c=0) \big\rangle_\mathbb{C}  \\
    B \equiv & \Im \big\langle d, \mu(\phi_c=0) \big\rangle_\mathbb{C} .
\end{align}
The integral yields modified Bessel function of the first kind
\begin{equation}
  I_{0}\left(\sqrt{A^2 + B^2}\right) = \frac{1}{2\pi} \int_{0}^{2\pi} d\phi \, e^{A c_{\phi} + B s_{\phi}} .
\end{equation}
Thus
\begin{align}
    \sqrt{A^2 + B^2} = & \sqrt{\Re \big\langle d, \mu(0) \big\rangle^2_\mathbb{C} + 
    \Im \big\langle d, \mu(\phi_c=0) \big\rangle^2_\mathbb{C}} \nonumber\\
    = & \left| \big\langle d, \mu(\phi_c=0) \big\rangle_\mathbb{C} \right| \nonumber\\
    = & \left|\ks_\mathbb{C}\right| ,
\end{align}
where $\ks_\mathbb{C}$ is calculated the same way as $\kappa$ (Eq.~\ref{eq:x}), except we use a complex inner product.
The $\phi_c$ marginalized likelihood becomes
\begin{equation}\label{eq:phase_marge}
\begin{split}
  \log &\mathcal{L}_\text{marg}^\phi = \log{\cal Z}_N - \frac{1}{2} \rho_\text{opt}^2 + \log I_0(|\ks_\mathbb{C}|) .
\end{split}
\end{equation}
We reiterate that this marginalized likelihood is valid only insofar as we trust our initial assumption, that the signal is dominated by $l=2,|m|=2$ modes.

\subsection{Distance marginalization}
In this subsection, we follow~\cite{BayesStar} (see also~\cite{Singer}) to derive a likelihood, which explicitly marginalizes over luminosity distance $D_L$.
Given a waveform at some reference distance $\mu(D_0)$, the waveform at an arbitrary distance is obtained by multiplication of a scale factor
\begin{equation}
	\mu_{j}(D_{L}) = \mu_{j}(D_{0})\, \left( \frac{D_{0}}{D_L} \right).
\end{equation}
As before, it is understood that $\mu$ is a function of whatever parameters are not explicitly marginalizing over.
Unlike time and phase, distance affects $\rho_{\text{opt}}$ in addition to $\ks$ (Eq.~\ref{eq:x}),
\begin{equation}\label{eq:rhoD}
\begin{split}
\ks(D_{L}) &= \ks(D_{0})\, \left( \frac{D_{0}}{D_L} \right), \\
\rho_{\text{opt}}^2(D_{L}) &= \rho_{\text{opt}}^2(D_{0})\, \left( \frac{D_{0}}{D_L} \right)^2 .
\end{split}
\end{equation}
Note that $\ks$ and $\rho_\text{opt}$ are implicit functions of whatever parameters we are not explicitly marginalizing over.

At a fixed distance, the likelihood is
\begin{equation}
\log \mathcal{L}(D_{L}) = \log{\cal Z}_N + \ks(D_{L}) - \frac{1}{2} \rho_\text{opt}^2(D_{L}),
\end{equation}
and the likelihood marginalized over luminosity distance is
\begin{equation}
\begin{split}
\log \mathcal{L}_\text{marg}^D = & \log{\cal Z}_N + \log {\cal L}_D ,
\end{split}
\end{equation}
where
\begin{align}
{\cal L}_D(\ks,\rho_\text{opt}) \equiv \int d D_{L}\, e^{\ks(D_{L}) - \frac{1}{2} \rho_\text{opt}^2(D_{L})} \pi(D_{L}) .
\end{align}
This integral to calculate $\log {\cal L}_D$ can be evaluated numerically.
This explicitly marginalized form is generally true for all gravitational-waves sources.
Its validity is only limited by the resolution of the numerical integral, though, cosmological redshifts adds additional complications, which we discuss in the next subsection.
One can construct a pre-computed lookup table $\log{\cal L}_D(\rho_\text{mf}, \rho_\text{opt})$ to facilitate fast and precise evaluation.

\subsection{Distance marginalization with cosmological effects}
There is a caveat for our discussion of distance marginalization in the previous subsection: when considering events at cosmological distances, the prior distributions for lab-frame masses become covariant with luminosity distance $D_L$ due to cosmological redshift.
A signal emitted with source-frame mass $m_s$ is observed with lab-frame mass given by
\begin{align}
m_l = (1+z) m_s .
\end{align}
In this subsection, ``mass'' $m$ is shorthand for an array of both primary and secondary mass.

Now we derive an expression for ${\cal L}_\text{marg}^D$, which can be applied to cosmological distances.
We start by specifying the prior on redshift and source-frame mass\footnote{Many previous analyses have assumed that this distribution is separable, however this marginalization technique does not require this.}:
\begin{align}
\pi(z, m_s) = \pi(z)\pi(m_s) .
\end{align}
Both $\pi(z)$ and $\pi(m_s)$ can be chosen using astrophysically motivated priors; see~e.g.,~\citep{mass,mass_uc,Fishbach}.
Whatever priors we choose for $\pi(z)$ and $\pi(m_s)$, they imply some prior for the lab-frame mass:
\begin{align}
\pi(z, m_l) = & \pi\big(z, m_l/(1+z)\big) 
\left|\frac{dm_s}{dm_l}\right| \nonumber\\
= & (1+z)^{-1} \pi\big(z, m_l/(1+z)\big) .
\end{align}

Now that we have converted the source-frame prior into a lab-frame prior, we can write down the distance-marginalized (redshift-marginalized) likelihood in terms of lab-frame quantities:
\begin{align}\label{eq:zsum}
{\cal L}_\text{marge}^z(\ks,\rho_\text{opt}) = \int dz \, 
{\cal L}(\ks,\rho_\text{opt}, z)
\pi(z|m_l) ,
\end{align}
where 
\begin{align}
{\cal L}(\ks,\rho_\text{opt}, z) = & {\cal Z}_N\, e^{\ks\left(D_L(z)\right) - \frac{1}{2}\rho_\text{opt}^2\left(D_L(z)\right)}.
\end{align}
Note that $\ks$ and $\rho_\text{opt}$ are implicit functions of whatever parameters we are not explicitly marginalizing over.

By creating a grid of $z$, we can create a look-up table for ${\cal L}(\ks,\rho_\text{opt},z)$, which allows for rapid evaluation of Eq.~\ref{eq:zsum}.
However, this means we will also need to create a look-up table for $\pi(z|m_l)$.
In order to derive this look-up table, we rewrite the joint prior on redshift and lab-frame mass can be rewritten like so
\begin{align}
\pi(z, m_l) = \pi(z|m_l) \pi(m_l) .
\end{align}
The marginalized lab-mass prior is 
\begin{align}
\pi(m_l) \equiv \int dz \, \pi(z, m_l) ,
\end{align}
which can be calculated numerically.
(We also need this distribution to provide to the sampler.)
Thus, the conditional prior we need for our look-up table is:
\begin{align}
\pi(z | m_l) = \pi(z, m_l) / \pi(m_l) .
\end{align}
With look-up tables for ${\cal L}(\ks, \rho_\text{opt}, z)$ and $\pi(z|m_l)$, the sampler can quickly evaluate ${\cal L}_\text{marge}^z$ by summing over the grid of $z$:
\begin{align}
{\cal L}_\text{marg}^z(\ks,\rho_\text{opt}) = \Delta z \sum_k {\cal L}(\ks, \rho_\text{opt}, z_k)
\pi(z_k| m_l) ,
\end{align}
where $\Delta z$ is the spacing of the redshift grid.
This allows us to carry out explicit distance marginalization while taking into account cosmological redshift.

\subsection{Marginalization with multiple parameters}
One must take care with the order of operations when implementing these marginalization schemes simultaneously.
We describe how to combine the three marginalization techniques described above.
The correct procedure is to start with Eq.~\ref{eq:phase_marge} and then marginalize over distance.
\begin{equation}
\begin{split}
  \log \mathcal{L}_\text{marg}^{\phi,D} = & \log{\cal Z}_N \\
  &+ 
  \log \int dD_{L}
  e^{I_0(|\ks_\mathbb{C}(D_L)|)
  -\frac{1}{2} \rho_\text{opt}^2(D_L)} \pi(dD_L) .
\end{split}
\end{equation}
Carrying out this integral numerically, one obtains a look-up table $\log{\cal L}_\text{marge}^{\phi,D}(\ks_\mathbb{C},\rho_\text{opt})$, which marginalizes over $\phi$ and $D_L$.
Finally, we add in $t$ marginalization by combining the look-up table with a fast Fourier transform
\begin{align}
{\cal L}_\text{marg}^{\phi, D, t}(\ks_\mathbb{C},\rho_\text{opt}) = 
\sum_k \pi_k \,
{\cal L}_\text{marg}^{\phi, D}\big(\ks_\mathbb{C}(k),\rho_\text{opt}(k)\big) .
\end{align}

\subsection{Reconstructing the unmarginalized posterior}
While explicitly marginalizing over parameters improves convergence and reduces runtime, the sampler will generate no posterior samples for the marginalized parameters.
Sometimes, we want posterior samples for these parameters.
In this subsection we explain how it is possible to generate them with an additional post-processing step.

The parameter we are most likely to be interested in reconstructing is the luminosity distance $D_L$.
Let us assume for the moment that this is the only parameter over which we have explicitly marginalized.
The first step to calculate the matched filter signal-to-noise ratio $\rho_\text{mf}$ and optimal signal-to-noise ratio $\rho_\text{opt}$ for each sample.
For one posterior sample $k$, the likelihood for distance is
\begin{equation}
  {\cal L}_k(d|D_{L}) = {\cal Z}_N \, 
  e^{\ks(\theta_k, D_{L}) - \frac{1}{2} \rho_\text{opt}^2(\theta_k, D_{L})} ,
\end{equation}
where $\ks(D_L)$ and $\rho_\text{opt}(D_L)$ are defined in Eq.~\ref{eq:rhoD}.
(When comparing with Eq.~\ref{eq:rhoD}, note that we have again made explicit the dependence on $\theta_k$ = whatever parameters we are not explicitly marginalizing over.)
Since this likelihood is one-dimensional, it is easy to calculate the posterior for sample $k$ using Bayes' theorem:
\begin{align}
p_k(D_L|d) = \frac{{\cal L}(d|D_{L}) \pi(D_L)}
{\int dL \, {\cal L}(d|D_{L}) \pi(D_L)} .
\end{align}
Using the posterior, one can construct a cumulative posterior distribution for sample $k$:
\begin{align}
P_k(D_L|d) = \int dD_L \, p_k(D_L|d) .
\end{align}
The integral can be carried out numerically.
The cumulative posterior distribution can be used to generate random values of $D_L$ for each posterior sample.
\begin{align}
D_L = P_k^{-1}({\tt rand})
\end{align}

Reconstructing the likelihood or posterior when multiple parameters have been explicitly marginalized over is more complicated.
However, one may use the following iterative algorithm.
\begin{enumerate}
\item For each sample $\theta_k$ marginalize over all originally marginalized parameters except one ($\lambda$).
\item Draw a single $\lambda$ sample from the marginalized likelihood times prior.
\item Add this $\lambda$ sample to the $\theta_k$ and return to step 1, this time not marginalizing over $\lambda$.
\end{enumerate}
Alternatively, one can skip the step of generating new samples in distance and calculate the likelihood of the data given $D_L$ marginalized over all other parameters,
\begin{align}
	{\cal L}(d|D_L) = &	\frac{1}{n}
    \sum_k^n {\cal L}_k(d|D_L) \nonumber\\
 = & \frac{{\cal Z}_N}{n}\sum_k^n e^{\ks(\theta_k, D_{L}) - \frac{1}{2} \rho_\text{opt}^2(\theta_k, D_{L})} .
\end{align}
This likelihood can be used in Eq.~\ref{eq:Ltot} to perform population inference on the distribution of source distances and/or redshifts.

\section{Posterior predictive distributions}\label{ppd}
\begin{figure}
\includegraphics[width=\linewidth]{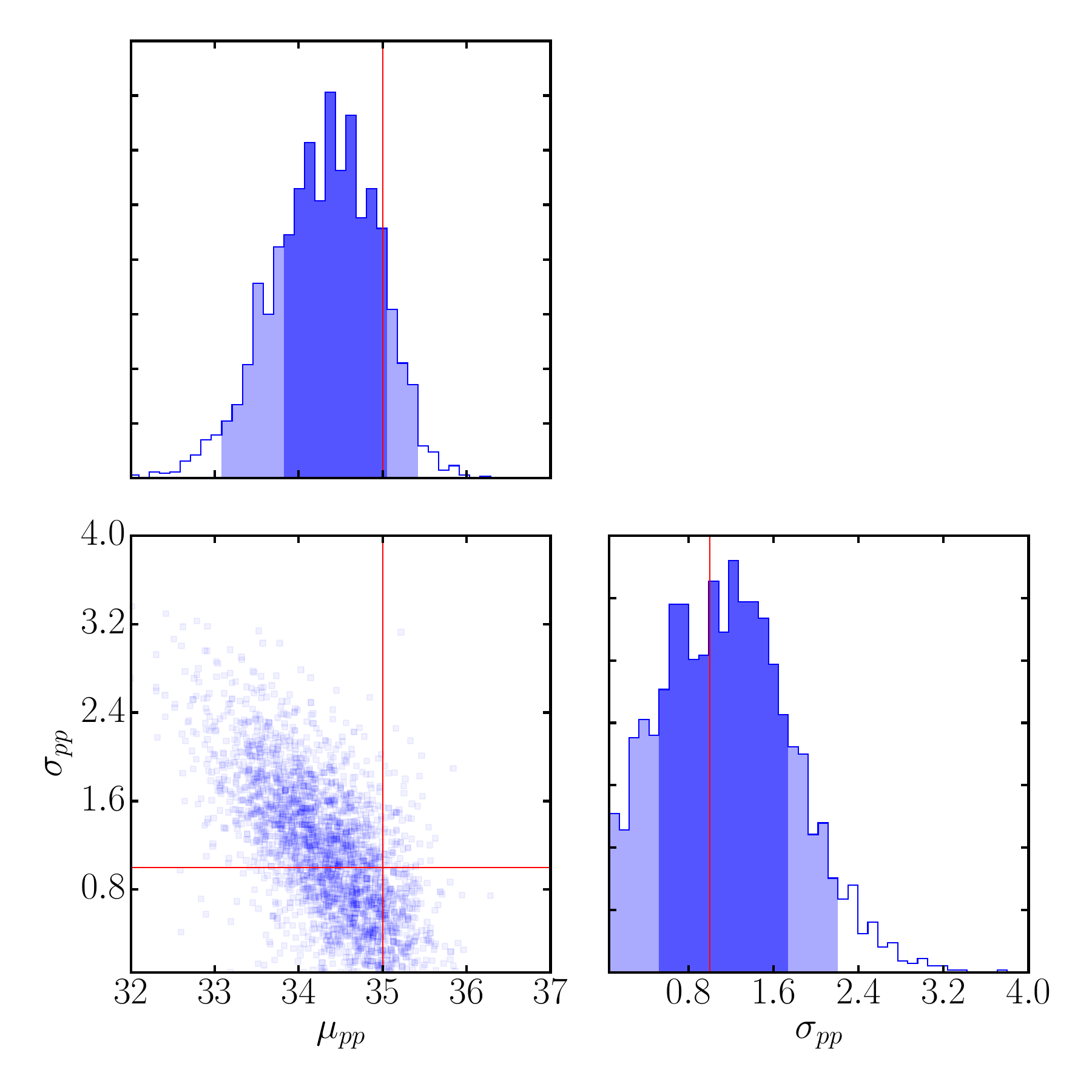}
\includegraphics[width=\linewidth]{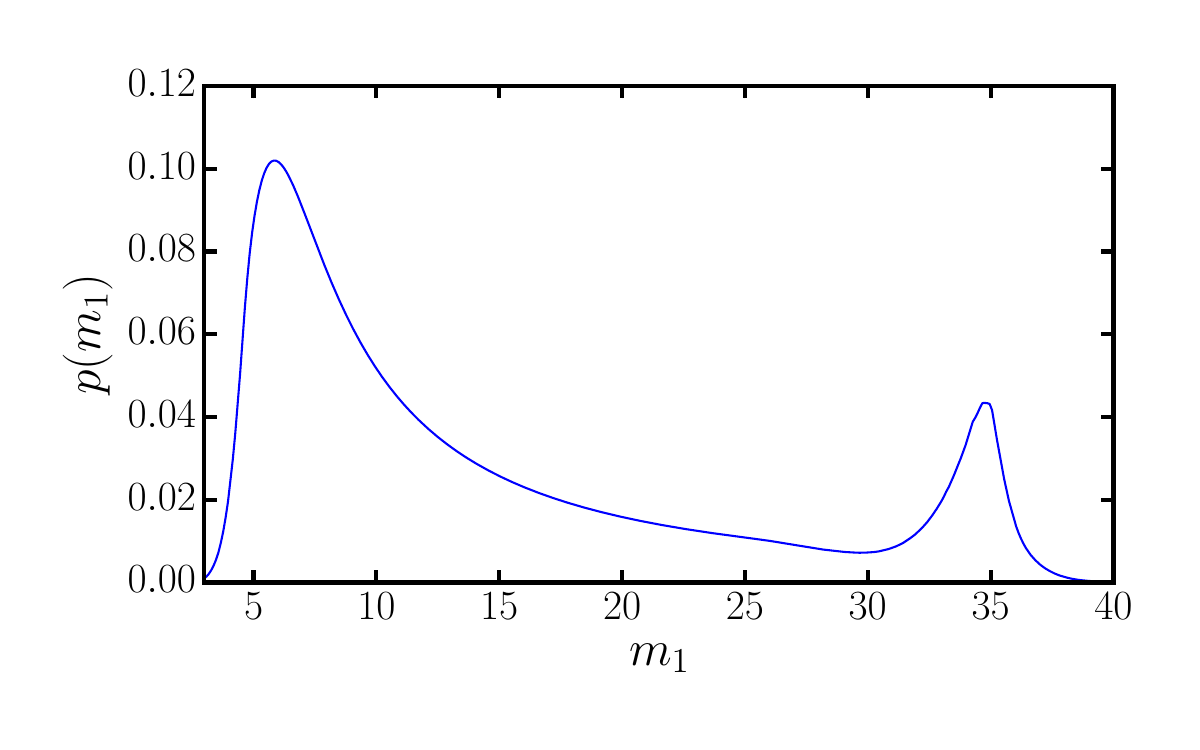}
\caption{
Top: an example corner plot from~\citep{mass} showing posteriors for hyper-parameters $\mu_\text{pp}$ and $\sigma_\text{pp}$.
Respectively, these two hyper-parameters describe the mean and width of a peak in the primary mass spectrum due to the presence of pulsational pair instability supernovae.
Bottom: an example of a posterior predictive distribution (PPD) for  primary black hole mass, calculated using the hyper-posterior distributions in the top panel (adapted from~\citep{mass}).
The PPD has a peak near $m_1=35$ because the hyper-posterior for $\mu_\text{pp}$ is maximal near this value.
The width of the PPD peak is consistent with the hyper-posterior for $\sigma_\text{pp}$.
\label{fig:ppd}
}
\end{figure}

The posterior predictive distribution (PPD) represents the updated prior on the parameters $\theta$ given the data $d$.
Recall that the hyper-posterior $p(\Lambda|d)$ describes our post-measurement knowledge of the hyper-parameters that describe the shape of the prior distribution $\pi(\theta)$.
The PPD answers the question: given this hyper-posterior, what does the distribution of $\pi(\theta)$ look like?
More precisely, it is the probability that the next event will have true parameter values $\theta$ given what we have learned about the population hyper-parameters $\Lambda$
\begin{align}
p_\Lambda(\theta|d) = \int d\Lambda \, p(\Lambda|d) 
\, \pi(\theta|\Lambda) .
\end{align}
The $\Lambda$ subscript helps us distinguish the PPD from the posterior $p(\theta|d)$.
The hyper-posterior sample version is
\begin{align}
p_\Lambda(\theta|d) = \frac{1}{n_s}\sum_k^{n_s} 
\pi(\theta|\Lambda_k) ,
\end{align}
where $k$ runs over $n_s$ hyper-posterior samples.
While the PPD is the best guess for what the distribution $\pi(\theta)$ looks like, it does not communicate information about the variability possible in $\pi(\theta)$ given uncertainty in $\Lambda$.
In order to convey this information, it can be useful to overplot many realizations of $\pi(\theta|\Lambda_k)$ where $\Lambda_k$ is a randomly selected hyper-posterior sample.
An example of a PPD is included in Fig.~\ref{fig:ppd}.

\section{Selection Effects}\label{selection}
In this section, we discuss how to carry out inference while taking into account selection effects, which arise from the fact that some events are easier to detect than others.
We loosely follow the arguments from~\cite{1606.04856}; however, see also~\cite{Mandel,Fishbach}.

Some gravitational-wave events are easier to detect than others.
All else equal, it is easier to detect binaries if they are closer, higher mass (at least, up until the point that they start to go out of the observing band), and with face-on/off inclination angles.
More subtle selection effects arise due to black hole spin~\citep[see, e.g.,][]{Ng}.
Typically, a gravitational-wave event is said to have been detected if it is observed with a matched-filter signal-to-noise ratio---maximized over extrinsic parameters $\theta_\text{extrinsic}$---above some threshold $\rho_\text{th}$
\begin{align}
    \rho_\text{mf}' \equiv
    \max_{\theta_\text{extrinsic}} \left(\rho_\text{mf}\right) > \rho_\text{th} .
\end{align}
Usually, $\rho_\text{th}=8$ for a single detector or $\rho_\text{th}=12$ for a $\geq2$ detector network.

\begin{figure}
    \centering
    \includegraphics[width=\linewidth]{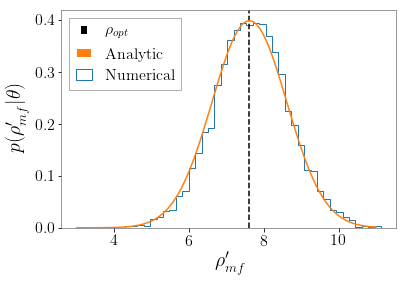}
    \caption{
    The distribution of matched filter signal-to-noise ratio maximized over phase for the same template in many noise realisations (blue). 
    The distribution peaks at $\rho_\text{opt}=7.6$ (dashed black).
    The theoretical distribution (Eq.~\ref{eq:rhomf}) is shown in orange.}
    \label{fig:selection}
\end{figure}

Selection effects are characterised by $p_\text{det}$, the probability that a signal exceeds the detection threshold.
There are different ways to calculate $p_\text{det}$ in practice.
The probability density function for $\rho_\text{mf}$ given $\theta$---the distribution of $\rho_\text{mf}$ arising from random noise fluctuations---is a normal distribution with mean $\rho_\text{opt}$ and unit variance
\begin{align}\label{eq:rhomf}
    p(\rho_\text{mf}'|\theta) = \frac{1}{\sqrt{2\pi}}
    \exp\left(-\frac{1}{2}\Big(\rho_\text{mf}'-\rho_\text{opt}(\theta)\Big)^2\right) ,
\end{align}
see Fig.~\ref{fig:selection}.
Thus,
\begin{align}
    p_\text{det}(\theta) = &
    \int_{\rho_\text{th}}^\infty dx \frac{1}{\sqrt{2\pi}} 
    \exp\left({-\frac{1}{2}\Big(x-\rho_\text{opt}(\theta)\Big)^2}\right) 
    \\
    = & 
    \frac{1}{2}\,\text{erfc}\left(\frac{\rho_\text{th}-\rho_\text{opt}(\theta)}{\sqrt{2}}\right) .
\end{align}
Alternatively, one may express $p_\text{det}$ as the ratio of the ``visible volume'' ${\cal V}(\theta)$ to the total spacetime volume ${\cal V}_\text{tot}$
\begin{align}\label{eq:Vtheta}
    p_\text{det}(\theta) = \frac{{\cal V}(\theta)}{{\cal V}_\text{tot}} .
\end{align}
The visible volume is typically calculated numerically with injected signals.

Given a population of $N$ events,
\begin{align}
{\cal L}(d,N|\Lambda,\text{det}) =
        \frac{1}{p_\text{det}(\Lambda|N)} 
        {\cal L}(d,N|\Lambda,R) .
\end{align}
In analogy to Eq.~\ref{eq:Vtheta}, the $p_\text{det}$ normalization factor can be calculated using the visible volume as a function of the hyper-parameters $\Lambda$ 
\begin{align}
    {\cal V}(\Lambda) \equiv \int d\theta
    {\cal V}(\Lambda) \pi(\theta|\Lambda) .
\end{align}
Naively, one might expect that
\begin{align}
    p_\text{det}(\Lambda|N) = \left(\frac{{\cal V}(\Lambda)}{{\cal V}_\text{tot}}\right)^N , 
\end{align}
but this expression is incorrect because it does not marginalize over the Poisson-distributed rate, which ends up changing the answer.
Marginalizing over the rate, we obtain
\begin{align}
    p_\text{det}(\Lambda|N) = &
    \int dR
    \left(\frac{{\cal V}(\Lambda)}{{\cal V}_\text{tot}}\right)^N
    \pi(N|R) \pi(R) \nonumber\\
    = & \int dR
    \left(\frac{{\cal V}(\Lambda)}{{\cal V}_\text{tot}}\right)^N
    \left[e^{-R {\cal V}(\Lambda)} \frac{{\cal V}(\Lambda)^N R^N}{N!} \right] \pi(R) \nonumber\\
    = & \left(\frac{{\cal V}(\Lambda)}{{\cal V}_\text{tot}}\right)^N \left[ \int dR \, e^{-R {\cal V}(\Lambda)} \frac{{\cal V}(\Lambda)^N R^N}{N!} \pi(R) \right] .
\end{align}
Note that $p_\text{det}$ depends on our prior for the rate $R$.
If we choose a uniform-in-log prior $\pi(R)\propto1/R$, we obtain 
\begin{align}
    p_\text{det}(\Lambda|N) \propto & \left(\frac{{\cal V}(\Lambda)}{{\cal V}_\text{tot}}\right)^N ,
\end{align}
which reproduces the results from~\cite{r_and_p}.
Note that 
\begin{align}
    {\cal L}(d|\Lambda,\text{det}) \neq \int d\theta
    {\cal L}(d|\theta,\text{det}) \, \pi(\theta | \Lambda) .
\end{align}

\section{Erratum}\label{erratum}
\begin{enumerate}
    \item In the original version of this manuscript, we included a subsection, ``Selection effects with a single event,'' which does not appear in this update.
    This section included formulas with errors including Eq.~89 and Eq.~95 of the arxiv version (Eq.~E2 and Eq.~E8 in the version published in PASA).
    Moreover, the section included a conceptual error since the idea of selection effects for single events does not make sense.
    Selection effects are intrinsically related to population studies, so they simply do not affect the analysis of single detections.
    It is interesting to consider how this comes about mathematically.
    While the single-event $\text{det}$ likelihood gains a factor of $p_\text{det}^{-1}$ (as correctly noted in the original manuscript), the single-event $\text{det}$ prior picks up a compensating factor of $p_\text{det}$, because the prior for detected events is not the same as the original (no $\text{det}$) prior.
    Since the $\text{det}$ posterior is proportional to the product of the likelihood and the prior, these two factors cancel, giving the original (no $\text{det}$) likelihood.
    \item In eight places in the original manuscript we referred to ``the odds ratio.''
    However, we should have referred simply to ``the odds.''
    In statistics, the odds refers to a ratio of probabilities.
    When we multiply the Bayes factor by the prior odds, we obtain the posterior odds.
    The odds ratio, which is also a statistical term, refers to a ratio of ratios.
    The present manuscript has been updated to fix this mistake.
\end{enumerate}

\end{appendix}

\bibliographystyle{pasa-mnras}
\bibliography{1r_lamboo_notes}

\end{document}